\documentclass{article}
\usepackage{graphics}
\def\ket#1{|\,#1\,\rangle}
\begin{document}
\title{\bf PPLN Waveguide for Quantum Communication}
\author{\bf S. Tanzilli\footnote{\it e-mail: tanzilli@unice.fr} $^{1}$, W. Tittel$^{2}$, H. De Riedmatten$^{2}$, H. Zbinden$^{2}$, P. Baldi$^{1}$,\\
\bf M. De Micheli$^{1}$, D.B. Ostrowsky$^{1}$ and N. Gisin$^{2}$\\
\\
$^1$ Laboratoire de Physique de la Mati\`{e}re Condens\'{e}e, CNRS
UMR 6622,\\ Universit\'e de Nice--Sophia Antipolis, Parc Valrose,
06108 Nice
Cedex 2, France\\
$^2$ Group of Applied Physics, University of Geneva,\\ 20,
rue de l`Ecole-de-M\'{e}decine, 1211 Geneva 4, Switzerland\\
}
\date{}
 \maketitle
\begin{center}
\underline{keywords}: PPLN waveguide, high efficiency, energy-time and time-bin entanglement
\end{center}
\begin{abstract}
We report on energy-time and time-bin entangled photon-pair sources based on a periodically poled lithium niobate (PPLN)
waveguide. Degenerate twin photons at 1314 nm wavelength are created by spontaneous parametric down-conversion and
coupled into standard telecom fibers. Our PPLN waveguide features a very high conversion efficiency of about $10^{-6}$,
roughly 4 orders of magnitude more than that obtained employing bulk crystals \cite{Tanzilli01a}. Even if using low
power laser diodes, this engenders a significant probability for creating two pairs at a time -- an important advantage
for some quantum communication protocols. We point out a simple means to characterize the pair creation probability in
case of a pulsed pump. To investigate the quality of the entangled states, we perform photon-pair interference
experiments, leading to visibilities of 97\% for the case of energy-time entanglement and of 84\% for the case of
time-bin entanglement. Although the last figure must still be improved, these tests demonstrate the high potential of
PPLN waveguide based sources to become a key element for future quantum communication schemes.
\end{abstract}
\newpage
\section{Introduction} \label{intro}

Sources creating entangled photon-pairs are an essential tool for
a variety of fundamental quantum optical experiments like tests of
Bell-inequalities
\cite{Bell64a,Mandel88a,Shih88a,Brendel92a,Kwiat93a,Weihs98a,Tittel98a,Kwiat99a},
quantum teleportation \cite{Bouwmeester97a,Boschi98a,Kim01a} and
entanglement swapping \cite{Jennewein01a}, as well as for more
applied fields of research such as quantum cryptography
\cite{Jennewein00a,Naik00a,Tittel00a} and "quantum" metrology
\cite{Migdall99a,Gisin98a}.

To date, the creation of entangled photon-pairs is usually assured
by spontaneous parametric down conversion (PDC) in non-linear bulk
crystals. However, despite enormous success in the above-mentioned
experiments, these sources suffer from low pair production
efficiencies. This turns out to be an important limitation,
especially if simultaneous creation of two pairs is required as in
quantum teleportation \cite{Bouwmeester97a}, entanglement swapping
\cite{Jennewein01a}, or in present schemes for the realization of
GHZ states \cite{Bouwmeester99a,Pan00a,Pan01a}. Taking advantage
of integrated optics, we recently built a new photon-pair source
based on a quasi-phase-matched optical waveguide implemented on
periodically poled lithium niobate (PPLN)\footnote{See also
related work by Sanaka \cite{Sanaka01a}}. We previously
demonstrated an improvement of the PDC efficiency by 4 orders of
magnitude compared to the highest efficiencies reported for bulk
crystals \cite{Tanzilli01a}. Here, we investigate the quality of
the created entanglement in terms of visibilities in two-photon
interference experiments for two different PPLN waveguide based
sources, the first one producing energy-time entangled
photon-pairs, and the second one generating time-bin entanglement.

The article is structured along the following lines: in Section
\ref{sec2}, we briefly discuss the advantage and the fabrication
of PPLN waveguides. In Section \ref{sec3}, we characterize the
efficiency of our source for photon-pair generation by means of
coincidence counting experiments based both on a CW and a pulsed
pump. The latter turns out to provide a simple means for
estimating the probability of creating a photon-pair per pump
pulse. Sections \ref{sec4} and \ref{sec5} then focus on two
different sources creating energy-time entangled as well as
time-bin entangled photon-pairs. We investigate the quality of the
entanglement via two-photon interference experiments. Finally, we
briefly conclude in Section \ref{sec6}.

\section{The PPLN waveguide}
\label{sec2}

The interaction of a pump field with a $\chi^{(2)}$ non-linear
medium leads, with a small probability, to the conversion of a
pump photon into so-called signal and idler photons. Naturally,
this process, known as spontaneous parametric down conversion, is
subjected to conservation of energy
\begin{equation}
\omega_p=\omega_s+\omega_i \label{EN}
\end{equation}
and momentum
\begin{equation}
\vec{k_p}=\vec{k_s}+\vec{k_i} \label{PM}
\end{equation}
where the indices label the frequency and k-vector of the pump,
signal and idler fields respectively. The latter equation is also
known as phase-matching condition and can be achieved, despite
chromatic dispersion, by employing different polarizations for the
three fields (so-called birefringent phase-matching).

\begin{figure}
\begin{center}
\resizebox{0.6\columnwidth}{!}{%
  \includegraphics{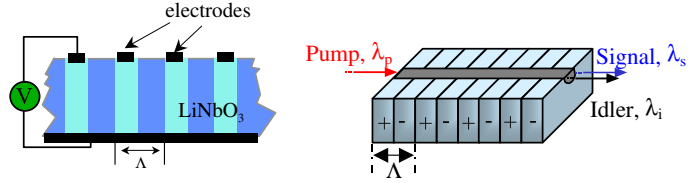}
} \caption{Poling of the lithium niobate substrate (left hand picture) and schematics of a PPLN waveguide (right hand
picture). $\Lambda$ denotes the poling period, $\lambda_p$, $\lambda_s$, $\lambda_i$ are the wavelength for the pump,
signal and idler photons, and + and - indicate the sign of the $\chi^{(2)}$ coefficient.} \label{fig1}
\end{center}
\end{figure}

In bulk optical configurations, the pump field is often slightly
focused into the non-linear crystal and the down-converted photons
are then collected into optical fibers by appropriate optics,
selecting photons created near the focal point of the first lens.
In opposition, the use of a guiding structure permits the
confinement of the pump, signal and idler beams over the entire
length of the waveguide (a few cm in our case). Therefore, it
clearly enables much higher down-conversion efficiencies. However,
natural birefringent phase-matching (Eq. \ref{PM}) is quite
difficult to obtain in a waveguiding structure. A better solution
is to employ so-called quasi-phase-matching (QPM) and to integrate
the waveguide on a periodically poled lithium niobate (PPLN)
substrate, i.e. a substrate where the ferroelectric polarization
of the material is periodically inverted (see Fig. \ref{fig1}).
The resulting phase-matching condition is then given by
\begin{equation}
\vec{k_p}=\vec{k_s}+\vec{k_i}+\vec{K}\hspace{1cm}
\mid\vec{K}\mid=\frac{2\pi}{\Lambda} \label{QPM}
\end{equation}
where $\vec{K}$ represents the effective grating-type k-vector and
$\Lambda$ is the poling period. By an appropriate choice of the
poling period ($\Lambda$), one can quasi-phase-match practically
any desired non-linear interaction between the pump, signal and
idler fields. In addition to an increased efficiency due to the
long interaction length, this allows working with the highest
non-linear coefficient of lithium niobate ($d_{33} \approx$ 30
pm/V) which is approximately six times larger than that ($d_{31}$)
commonly used in birefringent phase-matching in bulk crystals.

The inversion of the ferroelectric domains is done by applying a
high electric field ($\approx$ 20 kV/mm) periodically on a 500
$\mu$m thick lithium niobate sample (see Fig. \ref{fig1}),
inducing a periodic flip of the $\chi^{(2)}$ sign.

Once the PPLN substrate is obtained, the waveguide integration can
be done. We use a new one-step proton exchange procedure, called
"soft proton exchange" (SPE) \cite{Chanvillard00a}, working with a
low-acidity mixture of benzoic acid and lithium benzoate. During
this process, some lithium atoms (a fraction $x$) are replaced by
protons as described by the following reaction:
\begin{equation}
LiNbO_3+xH^+ \rightleftharpoons H_xLi_{1-x}NbO_3 + xLi^+
\label{SPE}
\end{equation}
This replacement locally creates a positive index variation of
about 0.03 along the aperture of a suitable mask (see Fig.
\ref{fig1}). It is important to note that SPE preserves both the
non-linear coefficient of the substrate as well as the orientation
of the domains.

\begin{figure}[h]
\begin{center}
\resizebox{0.4\columnwidth}{!}{%
  \includegraphics{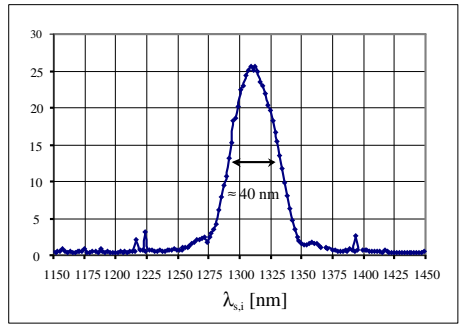}
} \caption{Measured spectrum for signal and idler photons in case of degenerate phase-matching.} \label{fig2}
\end{center}
\end{figure}

Our PPLN substrate features a length of 3.2 cm, and the waveguide
employed in the experiments has a 1/e width of 6 $\mu$m, a 1/e
depth of 2.1 $\mu$m and a poling period of 12.1 $\mu$m. Pumping
the sample at a wavelength of 657 nm, degenerate collinear
quasi-phase-matching is obtained for photons at a temperature of
100$^o$C. The temperature is high enough to avoid photorefractive
effects. Waveguide-losses at 1310 nm wavelength are smaller than
0.5 dB/cm. A typical spectrum of the parametric fluorescence in
the case of a coherent pump is shown in Fig. \ref{fig2}. The
degenerate down-converted twin-photons feature a bandwidth of
about 40 nm FWHM.
%\cleardoublepage

\section{Efficiency of the source}
\label{sec3}
\subsection{Coincidence Counting Experiment Using a CW Pump}

As reported recently \cite{Tanzilli01a}, our new source shows a
very high conversion efficiency. In the following, we briefly
present the experimental setup and review the major results.

\begin{figure}[h]
\begin{center}
\resizebox{0.9\textwidth}{!}{%
  \includegraphics{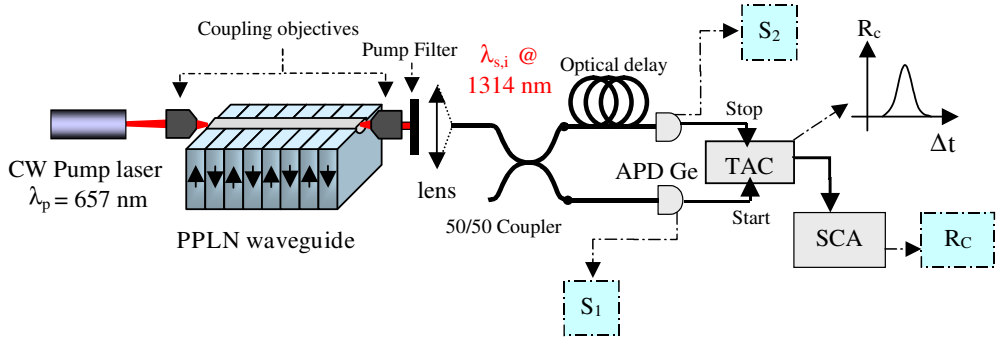}
} \caption{Schematics to characterize the downconversion efficiency of our PPLN waveguide.} \label{fig3}
\end{center}
\end{figure}

As shown in Fig. \ref{fig3}, the waveguide is pumped by a CW
laser. After absorption of the remaining pump photons by a filter,
signal and idler are separated using a 50/50 single mode fiber
optical beam-splitter. The photons are directed to passively
quenched LN$_{2}$ cooled germanium avalanche photodiodes
(Ge-APDs), operated in Geiger mode and featuring quantum
efficiencies of about 10\%. Net single count rates are denoted by
$S_1$ and $S_2$, respectively. The output from the APDs provide
the start and stop signals for a time to amplitude converter (TAC)
that records a coincidence histogram. A single channel analyzer
(SCA) enables counting the coincidence rate $R_C$ in a given
time-window.

As shown in \cite{Tanzilli01a}, the conversion efficiency $\eta$
can be expressed in terms of the experimental figures $S_1$, $S_2$
and $R_C$ (all are net count rates after subtraction of noise),
the pump power $P_P$, the pump wavelength $\lambda_P$ and the
usual constants h (Planck) and c (speed of light).
\begin{equation}
\eta=\frac{N}{N_P}=\frac{S_1 S_2}{2 R_C}\frac{h c }{P_P\lambda_P}
\label{efficiency}
\end{equation}

\noindent $N$ and $N_P$ denote the number of photon-pairs created
and the number of pump photons injected per second in the
waveguide, respectively. The factor of 2 in the denominator takes
account of the fact that only 50\% of the pairs are split at the
beam-splitter. Note that the calculation of the conversion
efficiency is based only on easily measurable quantities and does
not rely on only roughly known coupling constants and quantum
efficiencies of the detectors. With $P_P \approx 1 \mu$W measured
at the output of the waveguide, $\lambda_P$ = 657 nm, $S_i
\approx$ 150 kHz, and $R_C \approx$ 1500 Hz, we find a conversion
efficiency of about $2*10^{-6}$. This is at least 4 orders of
magnitude more than the efficiencies calculated for bulk sources
recently developed in Geneva \cite{Tittel98a}, in Los Alamos
\cite{Kwiat99a} or in Vienna \cite{Jennewein00a}. Note that $\eta$
is given for one spatial mode (defined by the single mode fibers)
but is not normalized to the spectral bandwidth.

\subsection{Coincidence Counting Experiment Using a Pulsed Pump}
\label{pulsedpump}

We now replace the continuous pump by a pulsed laser creating 400
ps pulses at a repetition rate of 80 MHz. Using a mean pump power
of only a few $\mu$W, the number of pump photons injected per
pulse in the waveguide is larger than $10^6$. Therefore, taking
into account the efficiency of the PPLN waveguide, the probability
to create two pairs within the same pump pulse is not negligible.
Obviously, the same holds for two pairs originating from two
subsequent pump pulses. Exact knowledge of this probability is
very important - the creation of two pairs within the same pump
pulse is at the heart of quantum teleportation
\cite{Bouwmeester97a}, entanglement swapping \cite{Pan01a} and
generation of GHZ states \cite{Bouwmeester99a,Pan00a} and
threatens the security of quantum cryptography systems based on
"single photons" produced by PDC \cite{Lutkenhaus00a}.

\begin{figure}[h]
\begin{center}
\resizebox{0.5\columnwidth}{!}{%
  \includegraphics{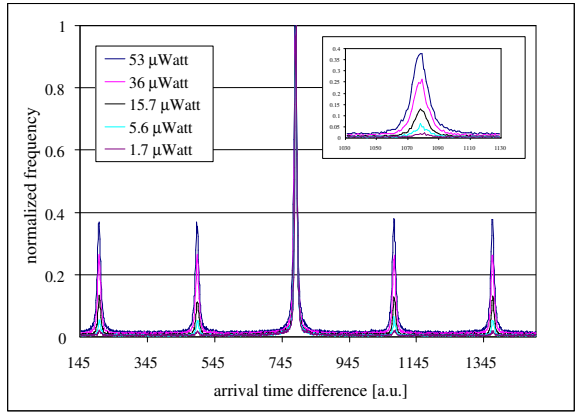}
} \caption{Central and satellite coincidence peaks as measured with a pulsed pump.} \label{fig4}
\end{center}
\end{figure}

Fig. \ref{fig4} shows typical coincidence histograms for different
mean pump power. In opposition to the CW case where only one
coincidence peak can be observed, we find a central peak
surrounded by equally spaced satellite peaks. The time difference
between neighboring peaks is 12.5 ns -- corresponding to the 80
MHz repetition rate of the pump laser --, and the ratio of the
respective magnitudes increases with the mean pump power.
\\

Here we will give only an intuitive explication for the emergence of these satellite peaks. A complete characterization
will be presented elsewhere. Since the probability to detect a down-converted photon is far from unity, we will for
instance find cases where the photon providing the start signal was created by a pump photon arriving at time $t_0$, and
where the stop-signal originates not from the simultaneously generated idler photon but from a photon generated by the
subsequent pump pulse. This coincidence will be located in the first coincidence satellite peak next to the central
peak. Since at least two photon-pairs have to be created in order to observe such an event, the ratio of the magnitude
of a satellite peak to central peak (where the creation of only one pair is sufficient) depends, in addition to the
detectors' quantum efficiency, on the probability to create a photon pair. It is thus possible to infer this very
important property from a simple coincidence measurement.

\section{Energy-Time Entanglement}
\label{sec4}

\begin{figure}[h]
\begin{center}
\resizebox{0.6\columnwidth}{!}{%
  \includegraphics{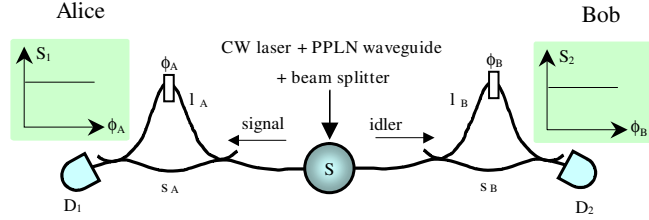}
} \caption{Schematics of a Franson-type setup to measure 2-photon interferences with energy-time entangled photons.}
\label{fig5}
\end{center}
\end{figure}

Apart from featuring a high down-conversion-efficiency, a two-photon source for quantum communication must be able to
create {\it{entangled}} photons. One possible way to infer the two photon quantum state would be to reconstruct its
density matrix \cite{White99a}. Another possibility is to perform a Bell-type experiment and to infer the degree of
entanglement via the measured two-photon interference visibility. The required setup for a test of Bell inequalities for
energy-time entangled photon pairs has been proposed by Franson in 1989 \cite{Franson89a} and first experiments have
been reported in 1992 \cite{Brendel92a,Kwiat93a}. A schematics is shown in Fig. \ref{fig5}. The source S is composed of
a CW laser diode with large coherence length ($\lambda_P$ = 657 nm), the PPLN waveguide to create photon-pairs and a
fiber optical 50/50 beam-splitter used to separate the twins. The photons are then sent to two analyzers at Alice's and
Bob's sides -- equally unbalanced Mach-Zehnder type interferometers made from standard telecommunication optical fibers
and Faraday mirrors (for a more complete characterization, see \cite{Tittel98a}). As in the coincidence counting
experiment, the detection is made by LN$_{2}$ cooled Ge-APDs D$_{1}$ and D$_{2}$. Note that two detectors are sufficient
to record the two-photon fringe visibility.

\begin{figure}[h]
\begin{center}
\resizebox{0.5\columnwidth}{!}{%
  \includegraphics{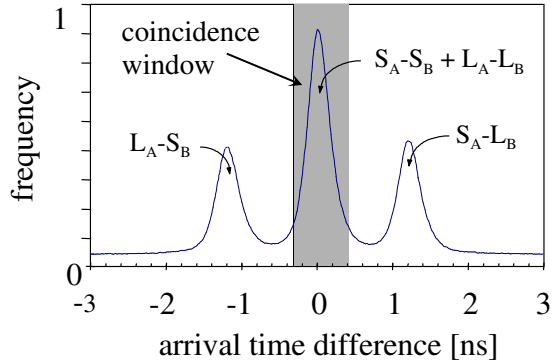}
} \caption{The coincidence histogram shows three different coincidence peaks, depending on the arms chosen in the
interferometers. The events where both photons pass through the same arms ($s_A-s_B$ and $l_A-l_B$) are
indistinguishable, leading to photon-pair interference. They can be discriminated from the non-interfering possibilities
$l_A-s_B$ and $s_A-l_B$ by means of a single channel analyzer (coincidence window). $s$ and $l$ denote the short and
long arms in (A)lice's and (B)ob's interferometers, respectively.} \label{fig6}
\end{center}
\end{figure}

As the arm length difference ($D_L$, about 20 cm of optical fiber)
is several orders of magnitude larger than the coherence length of
the single photons, no single photon interference is observed at
the outputs of the interferometers. The single count rates $S_1$
and $S_2$, respectively, are independent of the phases $\phi_A$
and $\phi_B$. However, if the coherence length of the pump laser
is larger than $D_L$, an "optical-path" entangled state can be
produced where either both down-converted photons pass through the
short arms or both through the long arms of the interferometers.
The non-interfering possibilities (the photons pass through
different arms) can be discarded using a high resolution
coincidence technique \cite{Brendel91a} (see Fig. \ref{fig6}),
enabling thus to observe quantum correlation with a visibility of
theoretically 100\%.

\begin{figure}[h]
\begin{center}
\resizebox{0.4\columnwidth}{!}{%
  \includegraphics{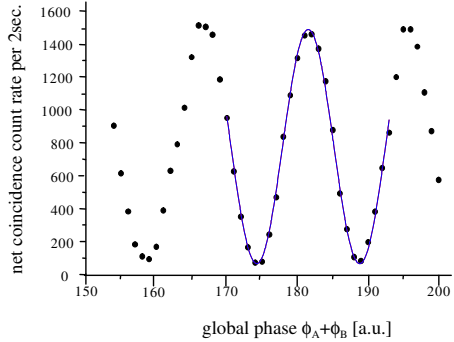}
} \caption{2-fold coincidence count rates as measured with energy-time entangled photons. The net visibility is of
97\%.} \label{fig7}
\end{center}
\end{figure}

Fig. \ref{fig7} shows the coincidence count rates per 2 sec. while varying the temperature (hence phase) in Alice's
interferometer. Although the single count rates do not show any interference, the coincidence count rates are described
by a sinusoidal function with a raw fringe visibility of (92$\pm$1)\%, and a net visibility (after subtraction of
accidental coincidences) of (97$\pm$1)\%. Note that the latter are {\it{not}} due to a reduced purity of the created
entanglement but to a combination of a high pair creation rate, losses, and small detector quantum efficiencies.
Therefore, in order to characterize the performance of the {\it{source}}, it is necessary to refer to the visibility
$V_{net}$ after subtraction of accidental coincidences. Since this visibility is close to the theoretical value of
100\%, we can conclude that the created state is indeed not far from a pure, maximally entangled state. In addition, we
can infer from $V_{raw}$ to a violation of Bell inequalities by 21 standard deviations. To do so, we have to assume that
the other three (not-recorded) coincidence count rates show a similar behaviour and that all coincidence count rates
depend only on the sum of the phases.

\section{Time-Bin Entanglement}
\label{sec5}

\begin{figure}[h]
\begin{center}
\resizebox{0.5\columnwidth}{!}{%
  \includegraphics{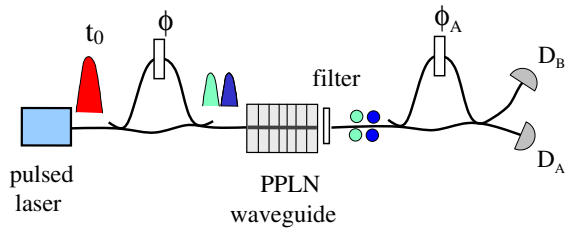}
} \caption{Experimental setup to create and measure time-bin entangled photon pairs.} \label{fig8}
\end{center}
\end{figure}

As already mentioned in the previous section, the coherence length of the pump photons has to be larger than the
path-length difference of the Mach-Zehnder interferometers in order to create an entangled state. Obviously, if we want
to pump our non-linear crystal with a pulsed laser, this coherence will be lost. The way to recreate the coherence to
the system is to place a third interferometer in the optical path of the pump photons as shown in Fig. \ref{fig8}.
\\A short light pulse emitted at a time $t_0$ enters the pump interferometer having a path-length difference greater than
the duration of the pulse. The pump pulse is then split into two pulses of smaller amplitudes following each other with
a fixed phase relation. Pumping now the PPLN waveguide, we create time-bin entangled photons in a state of the form
\begin{equation}
\ket{\psi}_{s,i}=\frac{1}{\sqrt2}\bigg(\ket{s}_s\ket{s}_i+
e^{i\phi}\ket{l}_s\ket{l}_i\bigg) \label{timebinstate}
\end{equation}
\noindent where $\ket{s}$ and $\ket{l}$ denote the short or long
arm of the first interferometer \cite{Brendel99b}. Depending on
the phase $\phi$, it is possible to obtain two out of the four
so-called Bell states. The two remaining Bell states can in
principle be created using optical switches and delay lines.
Finally, choosing different amplitudes for the long and short
components in the "pump" interferometer leads to the creation of
non-maximally entangled states.

The pulsed laser used in the experiment is the same as the one
used in the coincidence counting experiment discussed in Section
\ref{pulsedpump}. However, a grating reduces the spectral width
from 3 to 0.2 nm. Nevertheless, the pulse length remains 400 ps,
still longer than the one expected from the Fourier transformation
of the spectral width. The interferometer acting on the pump pulse
is made from single mode optical fibers at 655 nm wavelength, and
the transmission probabilities via the long and short arm are the
same in order to create maximally entangled states. The energy per
pump pulse is low enough to ensure that the creation of more than
one photon-pair at a time can be neglected. Finally, the photons
are analyzed in a way similar to the one described before.
However, for simplicity, both photons are sent into the same
interferometer and coincidences are recorded between the two
different output ports.

If we look at the coincidence events registered by a TAC placed
between $D_A$ and $D_B$, we find again three peaks denoting the
different paths taken by the signal and idler photons. However,
the $s_A-s_B$ and the $l_A-l_B$ events registered in the central
peak can now be due to photon-pairs created by a pump photon
passed trough either the short or the long arm of the pump
interferometer. Therefore the central peak includes four
possibilities: $s_P-s_A-s_B$, $l_P-s_A-s_B$, $s_P-l_A-l_B$ and
$l_P-l_A-l_B$. Since $s_P-s_A-s_B$ and $l_P-l_A-l_B$ are
distinguishable from the two other possibilities by the time
between emission of a pump pulse to detection at $D_A$ or $D_B$,
the coincidence visibility as observed with a SCA is limited to
50\%.

To regain the maximum visibility, one has to take into account the
emission time of the pump photon by recording three-fold
coincidences between the detection of the two down-converted
photons and the emission time of the pump pulse. In this way, only
the indistinguishable paths $l_P-s_A-s_B$ and $s_P-l_A-l_B$
remain, and the visibility can be increased to 100\%.

\begin{figure}[h]
\begin{center}
\resizebox{0.4\columnwidth}{!}{%
  \includegraphics{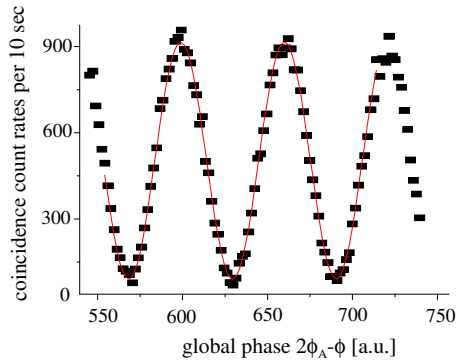}
} \caption{3-fold coincidence count rates as measured with time-bin entangled photons. The visibility is of 84\%.}
\label{fig9}
\end{center}
\end{figure}

Fig. \ref{fig9} shows the triple coincidence count rates per 10 sec. obtained while changing the phase in one
interferometer. The raw as well as the corrected net fringe visibilities are of around 84\%\footnote{Thanks to the
threefold coincidence required for each event, accidental coincidences can be neglected.}. This is still higher than the
limit of 71\% given by Bell inequalities, however, much lower than the maximally achievable value of 100\%. The reason
for the missing 16\% is not yet understood: it is not clear whether the reduced visibility is due to an alignment
problem in one of the interferometers, chromatic dispersion effects in the optical fibers or to a phenomenon related to
the PPLN waveguide.

\section{Conclusion and Outlook}
\label{sec6}

In this paper we investigated the performance of a
quasi-phase-matched PPLN waveguide for quantum communication. In a
CW coincidence counting experiment we first demonstrated a
conversion efficiency of 2*$10^{-6}$, corresponding to an
improvement of 4 orders of magnitude compared to bulk
configurations. This high efficiency engenders a significant
probability of generating more than one photon-pair at a time,
even if working with low power laser diodes. We pointed out a
simple means to deduce this probability from the magnitudes of
different coincidence count rates as observed in the case of a
pulsed pump. In addition to a high down-conversion efficiency, we
observed a net visibility of 97\% in a Franson-type test of
energy-time entangled photon-pairs, demonstrating the high purity
of entanglement. This result together with the high conversion
efficiency demonstrates the huge potential of PPLN waveguides for
future quantum communication sources. Finally, we observed a
two-photon visibility of 84\% in the case of time-bin entangled
pairs. The reason for the limited visibility is not yet
understood, but we hope to improve the performance of the source
and to make it a key element for future experiments where high
photon-pair production rates remain the critical point.

Efforts are underway to develop new PPLN waveguides enabling
down-conversion from 710 nm to 1310 and 1550 nm, or generating
signal and idler photons emitted in opposite directions. This
requires different poling periods than the one used for the
present work. Finally, a very interesting extension of the PPLN
waveguide is to build all-pigtailed devices, or devices where the
whole photon-pair source including interferometer and coupler to
separate the entangled photons is integrated on a single chip.
This could open the way to integrated quantum optics.

\textit{\\\\\\We acknowledge financial support by the Cost Action P2 "Application of non-linear optical phenomena" and
the ESF "Quantum Information Theory and Communications" as well as the Swiss FNRS and the IST-FET ``QuComm'' project of
the European Commission, partly financed by the Swiss OFES.}

%
%--------------------------------------------------------------
% BIBLIOGRAPHY
%--------------------------------------------------------------

%
\end{document}